\documentclass[oldversion]{aa}
\usepackage{graphicx,natbib}
\usepackage{txfonts}

\newcommand{\Deg}{\hbox{${}^\circ$}}
\newcommand{\Min}{\hbox{${}^{\prime}$}}
\newcommand{\Sec}{\hbox{${}^{\prime\prime}$}}

\begin{document}
   \title{$XMM-Newton$ observations of the hot spot galaxy NGC~2903}


   \author{D.~P\'erez-Ram\'{i}rez     \inst{ 1}
   \and    M.~D.~Caballero-Garc\'{i}a \inst{ 2}
   \and   J.~Ebrero                   \inst{ 3}
   \and   S.~Leon                     \inst{ 4}
 }

   \offprints{D.~P\'erez-Ram\'{i}rez}

   \institute{
   Universidad de Ja\'en, Campus Las Lagunillas, s/n, 23007, Spain
   \and  University of Cambridge, Institute of Astronomy, Cambridge CB3 0HA, UK                                    
   \and SRON-Netherlands Institute for Space Research, 3584 CA, Utrecht, The Netherlands                                                                 
   \and Joint Alma Observatory/ESO, Las Condes, Santiago, Chile   }

   \date{}

  \abstract{ 
We report on the first deeper X-ray broad-band observation of the hot
spot galaxy NGC~2903 obtained with \textsl{XMM-Newton}. X-ray imaging
and spectra of the spiral barred galaxy NGC~2903 were obtained from
\textsl{XMM-Newton} archival data to study its X-ray population and
the conditions of the hot gas in its central region. We investigate
the spectral properties of the discrete point-source population and
give estimates of their X-ray spectral parameters. By analysing the
RGS spectra, we derive temperature and abundances for the hot gas
located in its central region.  A total of six X-ray point sources
(four of them ULX candidates) were detected in the energy range of
0.3--10.0\,keV located within the galaxy $D_{25}$ optical disk. Three
out these sources are detected for the first time, and one of them,
XMM-NGC2903~X2 with a luminosity of higher than 10$^{39}$ erg
s$^{-1}$. After fitting three different models, we were able to
estimate their luminosities, which are compatible with those of
binaries with a compact object in the form of black holes (BHs) rather
than neutron stars (NSs). We extracted the combined first-order RGS1
and RGS2 spectra of its central region, which display several emission
lines. The spectrum is dominated by a strong O\,{\sc viii} Ly$\alpha$
emission line along with Ne\,{\sc x} Ly$\alpha$ and several Fe\,{\sc
xvii} features. The O\,{\sc vii} complex is also significantly
detected, although only the forbidden and resonance lines could be
resolved. Both O\,{\sc vii} $f$ and $r$ lines seem to be of similar
strength, which is consistent with the presence of the collisionally
ionized gas that is typical of starburst galaxies. We fitted the
spectrum to a model for a plasma in collisional ionization equilibrium
(CIE) and the continuum was modelled with a power law, resulting in a
plasma temperature of $T = 0.31 \pm 0.01$~keV and an emission measure
$EM \equiv n_Hn_eV =6.4_{-0.4}^{+0.5}\times 10^{61}$~cm$^{-3}$. We
also estimated abundances that are consistent with solar values.  }

  \keywords{galaxies: individual: \object{NGC~2903} --  galaxies: ISM -- 
galaxies: starburst -- X-rays: galaxies -- X-rays: binaries }

   \maketitle
%

\section{Introduction}

Starbursts galaxies are powerful sources of X-ray emission because of
their enhanced star formation (SF) activity. Galaxies in the starburst
stage appear to be a common phenomenon in the Universe on all distance
scales. They possibly share the same X-ray emission processes, hence
it is important to investigate these mechanisms in the local
systems. The study of X-ray emission from barred galaxies is of
particular interest because they are a class of spirals that appear to
frequently go through starburst episodes, and its possible association
with the knots of SF distributed along their bar and in their
circumnuclear regions (CNRs). Bars are often invoked as a possible
mechanism to fuel nuclear/circumnuclear SF \citep{hs94}. A significant
fraction of barred galaxies have nuclear rings within 1 kpc of the
nucleus of the galaxy, which are believed to be formed by gas
accumulation near galactic resonances. These nuclear rings are the
locations of strong density enhancements of gas about which enhanced
SF is often detected (\citealp{pst95,bc96,sh99,dkp00})

The different energetic phenomena detected in starburst galaxies at
high-energy gamma rays (e.g. NGC 253 at TeV by HESS, \citealp{ace09};
M82 at TeV by VERITAS, \citealp{acc09}; both galaxies at GeV by Fermi,
\citealp{abd10}; NGC 4945 and NGC 1068 at GeV by Fermi,
\citealp{len10}) are also manifested by their complex X-ray spectral
characteristics. \citet{pr02} quantitatively assessed the roles of the
various X-ray emission mechanisms and their main X-ray spectral
component at different energy ranges. At low energies ($\le$2\,keV),
the spectrum is dominated by the contribution of a diffuse plasma
(with low temperature, kT $\le$1\,keV), which originates mainly from
the shock heating in the interaction of the hot low-density wind with
the ambient high-density ISM \citep{ss00}, and, in a minor
contribution, from the galactic wind. At higher energies (2--10\,keV),
the spectrum seems to be dominated by emission from X-ray binaries
(XRBs; whose primaries are mainly neutron stars (NSs)) with a possible
contribution from either non-thermal Compton emission or an AGN
\citep{pr02}. If a bright AGN is not present, then the X-ray emission
from these galaxies is dominated by the contribution of a small number
of bright extranuclear point-like sources, which appear to be XRB
systems with black holes as primary components rather than NSs
(\citealp{f1}; \citealp{mc04}); in addition, they have X-ray
luminosities exceeding the Eddington luminosity of a $10\,{\rm
M}_{\odot}$ black hole, implying that they are being considered
ultraluminous X-ray sources (ULXs, \citealp{mak00}).

NGC~2903 is a large SAB(rs)bc starburst galaxy with a symmetric strong
bar that is considered typical of this class of galaxies
\citep{ls02}. Its proximity (8.9$\,Mpc$, \citealp{dk00}) and infrared
luminosity (in 8--1000$\mu$m $\sim$ 9.1$\times$10$^{9}$ $L\odot$;
\citealp{saka99}) has made it the subject of numerous studies at all
wavelengths, including radio (\citealp{tsa06,leon08}), infrared
(\citealp{sim88}), and optical (\citealp{bres05}). Early NIR studies
of this galaxy have shown that a considerable SF is occurring within
the complex hotspot morphology in the nuclear region
(e.g. \citealp{wb85,sim88}).

Previous X-ray studies of NGC~2903 have been based on $ROSAT$
observations (\citealp{thj03}). The soft X-ray band (0.1--2.4\,keV)
detects a galaxy with extended X-ray emission associated with enhanced
SF activity in both the disk and nuclear region, with a more
significant contribution from the disk. The nucleus and three ULXs
were previously reported within the $ROSAT$ energy band
(\citealp{lb05,rw00}). Moreover, weak emission outside the disk could
be discerned in the PSPC image suggesting the presence of a possible
halo (\citealp{thj03}).

In this paper, we report the analysis of the six sources (four
potential ULXs) located in NGC~2903 obtained with \textsl{XMM-Newton}
data, as well as the results of the RGS spectral analysis of its
central region.
 
The paper is organized as follows.  Section~\ref{sec:obs} describes
the observations and data reduction of the X-ray
observations. Detected sources and the main results of the
cross-identification search for their potential counterparts at other
wavelengths are given in
Section~\ref{sec:point}. Section~\ref{sec:prop} discusses the main
spectral properties of the X-ray point sources detected inside the
$D_{25}$ optical disk based on the fitting with main phenomenological
models. Results from the analysis of the RGS spectra of the central
region of NGC~2903 are presented in Section~\ref{sec:rgs}. We give the
main conclusions of our analysis in Section~\ref{sec:concl}.

\section{X-ray observations and data reduction}
\label{sec:obs}

\begin{table*}
 \begin{minipage}{190mm}
  \caption{Sources detected in the galaxy with {\it XMM-Newton}.}
  \label{log_obs}
  \begin{tabular}{@{}lcccc@{}}
  \hline
   Source name     &   R.A. ($h:m:s$)     &   Dec. ($^{\circ}:m:s$) & Position errors ($\Sec$) &    Rate (\,counts$/$s)    \\ 
\hline
 XMM-NGC2903~X-1           &  09$^{\rm h}$ 32$^{\rm m}$ 09\fs922 &  $+21$\Deg~30\Min~05.55\Sec & 0.3  & $0.1448{\pm}0.0018$  \\
 XMM-NGC2903~X-2           &  09$^{\rm h}$ 32$^{\rm m}$ 06\fs224 &  $+21$\Deg~30\Min~58.44\Sec & 0.2  & $0.0468{\pm}0.0011$   \\
 XMM-NGC2903~X-3           &  09$^{\rm h}$ 32$^{\rm m}$ 09\fs719 &  $+21$\Deg~31\Min~02.94\Sec & 1.5  & $0.0155{\pm}0.0007$ \\
 XMM-NGC2903~X-4           &  09$^{\rm h}$ 32$^{\rm m}$ 05\fs378 &  $+21$\Deg~32\Min~33.90\Sec & 0.3  & $0.0250{\pm}0.0008$  \\
 XMM-NGC2903~X-5           &  09$^{\rm h}$ 32$^{\rm m}$ 12\fs331 &  $+21$\Deg~29\Min~22.35\Sec & 0.5  & $0.0209{\pm}0.0008$ \\
 XMM-NGC2903~X-6           &  09$^{\rm h}$ 32$^{\rm m}$ 08\fs934 &  $+21$\Deg~29\Min~02.62\Sec & 1.6  & $0.0221{\pm}0.0008$ \\
\hline
\end{tabular}
\end{minipage}
\end{table*}

The dataset analysed was obtained from the \textsl{XMM-Newton} public
data archive. \textsl{XMM-Newton} observed NGC~2903 during its orbit
1715 (2009 April 21st; 96\,ksec on source time; Obs.ID: 0556280301;
PI: A. Diaz). All three EPIC cameras were operated in ``full frame''
mode with the thin optical filter.

The event lists were pipeline-processed and all data products created
using the {\small SAS} ({\it Science Analysis Software}) {\small
v10.0.0} and the EPIC calibration as of July 2010.

The EPIC-pn camera with a higher effective area than the EPIC-MOS
cameras was selected for the source detection and spectral extraction
in our analysis. We created a combined EPIC-pn image of NGC~2903 shown
in Fig~1. The image was adaptively smoothed and scaled for the optimal
visual presentation.
 
We applied the standard time and flare filtering for the observation.
A high energy full-field background light curve was produced for all
exposures to check for high background intervals of soft proton
flares. We used this light curve at energies above 10 keV to select
time intervals with low background (good time intervals, GTIs). Time
intervals with count rates $>$ 0.4\,ct s$^{-1}$ were cut from
subsequent data analysis\footnote{Information provided at
"node50.html" of the User Scientific Guide.}. Only those events
measured in regions away from the CCD borders or bad pixels (FLAG=0)
and single and double events for the PN camera ($PATTERN \le 4$) were
considered, as recommended for the PN camera.  We checked for pile-up
and found that this was not significant ($\le$5\% for the high energy
channels) for this observation. These corrections resulted in typical
exposure times of 47 ks.

Source detection was performed in several bands for the PN camera
simultaneously, using the psechain {\small SAS} task. This process
detected a total of six sources. The source identification used in
this work, the coordinates as given by the psechain task, the
point-source location accuracy (PSLA), the count rate for each source
in the 0.3-10keV energy range, and the intrinsic X-ray luminosity
within the same range considering a distance to the galaxy of
8.9$\,Mpc$ (\citealp{dk00}) are listed in Table~1. The pipeline also
created images, background images, and exposure maps for the
observation in different energy bands. In our analysis, we use the
energy range 0.3-10.0\,keV.

\section{X-ray point sources in the NGC~2903}
\label{sec:point}

We detected a total of six sources within the isophotal D$_{25}$
ellipse of the galaxy. We carried out a search for possible
counterparts to the X-ray point sources by cross-identifying the
location of the X-ray sources with specific catalogues for NGC~2903.
Although we obtained the respective PSLAs by means of the psechain
task for each source, we considered an absolute error of 2'', which
also accounts for the pointing uncertainties of $XMM-Newton$,
according to the latest calibration
report\footnote{http://xmm2.esac.esa.int/docs/documents/CAL-TN-0018.pdf}.

We cross-correlated the sources with the catalogues of \citet{thj03}
and \citet{ppz10} for the locations of H$_{II}$ regions and SF knots,
the work by \citet{sab09} with new detections of supernova remnants
(SNRs), and also the catalogue of \citet{al01} with the locations of
nuclear star clusters. We also cross-identified the location of the
X-ray sources with the ULX catalogues of \citet{lb05} and the $ROSAT$
high resolution survey of galaxies of \citet{rw00}. We did not find
any counterparts to our sources among the listed SNRs, nor the nuclear
star clusters.

Only three of the six X-ray sources have been previously reported,
namely XMM-NGC2903~X1, XMM-NGC2903~X3, and XMM-NGC2903~X4.  The source
XMM-NGC2903 X1, located at 3.68'' from the centre, was classified by
\citet{lb05} as an ULX, NGC~2903~X1, based on $ROSAT$ data at a
distance of 2.66'' from our source. Given the $ROSAT$ angular
resolution (2''), this source is within the position error of our
source. Eight radio sources are reported within 7'' of our position by
\citet{tsa06}, two of them at a separation of $\sim$ 1''. Several
$H_{II}$ regions are also found in the region by \citet{pcp97} two of
them being also at distances of less than 2''.

At a separation of 3.18'' from XMM-NGC2903~X3, NGC~2903~X-2 was
reported by \citet{rw00} based on $ROSAT$ high resolution images.

Finally for XMM-NGC2903~X4, an ULX source located at a separation of
3.63'', NGC 2903~X-1, was also reported by \citet{rw00}. A H$_{II}$
ionized region at a distance of 2.12'' was detected by \citet{thj03}
and \citet{ppz10}. In the SIMBAD database, we did not find any
counterpart or entry to XMM-NGC2903~X2, XMM-NGC2903~X5, nor to
XMM-NGC2903~X6, within a 10'' radius query.

In their ULX catalogue, \citet{lb05} reported two ULXs, namely X1,
which is coincident with our central source XMM-NGC2903~X1, and X2,
which is not detected in our observations.

   \begin{figure}
   \resizebox{\hsize}{!}{\includegraphics[angle=0]{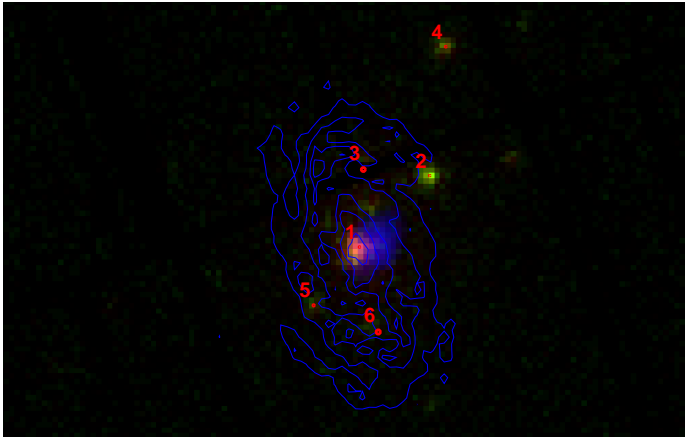}}
      \caption{Adaptively smoothed EPIC pn image of NGC~2903. The
        colours correspond to the energy bands, soft (0.5--2.0 keV,
        red), medium (2.0--4.5 keV, green), and hard (4.5--7.5 keV,
        blue). Overplotted in red, the location of the X-ray detected
        sources and in blue are the contours corresponding to the
        optical emission from this galaxy. North is upwards and east
        is leftwards.}
         \label{fig5}
   \end{figure}

\section{Spectral analysis of X-rays sources detected in NGC~2903}
\label{sec:prop}

We extracted the spectra for the sources from a circular region
centred on the source (with centroids shown in Table 1). The
background was extracted from a circular region not far from the
sources and away from the boundaries of the chip and the nucleus of
the galaxy. The EPIC spectra were grouped to contain a minimum of 20
counts per spectral bin to allow us to perform $\chi$$^{2}$
statistics, and fit to analytical models using the XSPEC 12.0 fitting
package \citep{a1}. All errors are given at the 90 percent confidence
level unless stated otherwise. The PN spectra were fitted in the
0.3--10.0\,keV band.

We fitted the spectra from all sources with the simplest
(non-composite) phenomenological models, i.e. multi-color disk
emission (diskbb; \citealp{mit84}) and power-law emission (best-fit
parameters shown in Table 2). These models have been proven to
successfully provide a phenomenological description of the
$XMM-Newton$ and $CHANDRA$ spectra of a sample of ULXs (e.g. in the
galaxies M51 and NGC 4490/85; \citealp{yoshida10,mak00})
. Furthermore, we applied the thermal bremsstrahlung {\tt bremss}
model. The obtained thermal temperatures were in the range 2-7\,keV,
indicative in some cases of high-energy curvature.  Soft-energy
emission from the diffuse galaxy was included in the model,
considering fixed metal (solar) abundances (mekal model in XSPEC). The
derived thermal temperature of the diffuse plasma is ${\rm kT}_{\rm
mekal}=0.5{\pm}0.2$ \,keV and the column density
${\sim}0.03{\times}10^{+22}\,{\rm cm}^{-2}$, closely in agreement with
the estimated column density $N_H = 2.91 \times 10^{20}$~cm$^{- 2}$
\citep{kal05} and indicative of low intrinsic column absorption to the
hosting galaxy. The X-ray luminosities obtained for these sources are
in the range $5{\times}10^{+38}- 10^{+39}\,erg\,s^{-1}$, with
XMM-NGC2903~X1 making it a strong low-luminosity AGN
candidate. Further $CHANDRA$ observations with higher spatial
resolution are required to confirm/discard this possibility.

XMM-NGC2903~X2, XMM-NGC2903~X3, and XMM-NGC2903~X4 reach and/or exceed
the Eddington luminosity of a stellar-mass black hole of $M=10\,{\rm
M}_{\odot}$ ($L_{\rm EDD}=1.3{\times}10^{38}M/{\rm M}_{\odot}$). For
XMM-NGC2903~X5, and XMM-NGC2903~X6, they do not reach this level and
emit as BH candidates in ther bright intermediate states
\citep{rm06}. For three of these sources, namely XMM-NGC2903~X1,
XMM-NGC2903~X5, and XMM-NGC2903~X6, the power-law model was a more
accurate description than the diskbb model. For XMM-NGC2903~X1, a
power-law solution is in good agreement with what we expect from a
low-signal AGN spectra and accreting BHs. It implies that non-thermal
processes play an important role in producing the spectra of these
sources. For the remaining sources, the breaks closely resemble those
previously observed in X-ray spectra from high-quality data from the
X-ray curved spectra of some ULXs (e.g., M82 X-1;
\citealp{miyawaki09}).

\begin{table*}
\begin{minipage}{190mm}
  \caption{Best-fit model spectra parameters for the sources detected in NGC 2903 with {\it XMM-Newton}.}
  \label{log_obs}
  \begin{tabular}{lcccccc}
  \hline
              &                &               &               &                 &               &          \\ 
Parameter     &  XMM-NGC2903   & XMM-NGC2903   & XMM-NGC2903   & XMM-NGC2903    & XMM-NGC2903   & XMM-NGC2903\\
              &       X-1      &    X-2        &        X-3    &    X-4         &    X-5        &     X-6\\
              &                &               &               &                 &               &          \\ 
  \hline  
              &                &               &               &                 &               &          \\ 
              &                &               &   P\footnote{The abbreviations for the models: ``P'' for phabs(mekal + powerlaw), ``D'' for phabs(mekal + diskbb) and ``B'' for phabs(mekal + bremss)}                        &                 &        &   \\
              &                &               &               &                 &               &          \\ 
  \hline 
              &                &               &               &                 &               &          \\ 
nH (10$^{22}$cm$^{-2}$)&  0.04$\pm$0.02 & 0.16$\pm$0.05 & 0.07$\pm$0.03 & 0.53$\pm$0.02  & 0.03$\pm$0.03 & 0.01$\pm$0.01\\
$\Gamma$      &  2.15$\pm$0.05 & 1.88$\pm$0.05 & 2.52$\pm$0.18 & 2.76$\pm$0.16  & 1.76$\pm$0.11 & 1.76$\pm$0.11\\
kT (keV)      &  0.59$\pm$0.02 & 0.29$\pm$0.12 & 0.53$\pm$0.14 & 0.14$\pm$0.02  & 0.47$\pm$0.18 & 0.54$\pm$0.15\\
$\chi$$^{2}$/$\nu$& 1.07(240.68/224)& 1.52(150.58/99)& 0.64(23.08/36)& 0.69(37.18/54)&  0.94(43.20/46)& 1.06(50.91/48) \\
F$^{unabs}_{0.3-10}$(ergs/cm$^{2}$/s)(10$^{-13}$)&3.4898$\pm$0.0123 & 2.2282$\pm$0.0101 & 1.6114$\pm$0.3893 & 0.7095$\pm$2.6372 & 0.6427$\pm$0.1143& 0.6006 $\pm$0.0140  \\
Log(Luminosity)& $39.52(^{+0.03}_{-0.04})$ & $39.32(^{+0.06}_{-0.06})$& $39.18(^{+0.22}_{-0.07})$ & $39.83(^{+0.03}_{-0.73})$&$38.78(^{+0.07}_{-0.09})$&$38.76(^{+0.06}_{-0.08})$\\
& & & & & & \\
\hline
              &                &               &               &                 &               &          \\ 
              &                &               &    D          &                &                &          \\ 
\hline
              &                &               &               &                 &               &          \\ 
nH (10$^{22}$cm$^{-2}$)    &0.005$\pm$0.005 &0.006$\pm$0.006&0.008$\pm$0.008&0.08$\pm$0.03   &0.006$\pm$0.006&0.005$\pm$0.005\\
kT$_{in}$ (keV)   & 0.66$\pm$0.05  & 1.39$\pm$0.08 & 0.47$\pm$0.07 &0.95$\pm$0.11   &1.4$\pm$0.4    &1.4$\pm$0.3\\
kT$_{mekal}$ (keV)&0.52$\pm$0.02   & 0.43$\pm$0.13 & 0.49$\pm$0.16 &0.7$\pm$0.5     &0.31$\pm$0.03  &0.30 $\pm$0.03\\
$\chi$$^{2}$/$\nu$& 1.84(412.29/224)& 1.46(144.77/99)& 0.90(32.33/36)& 0.66(35.48/54) & 1.34(61.83/46) & 1.48(71.02/48)\\
F$^{unabs}_{0.3-10}$(ergs/cm$^{2}$/s)(10$^{-13}$)&2.4453$\pm$0.0457&1.4862$\pm$0.0249&0.9635$\pm$0.0533&0.8101$\pm$0.6931&0.5474$\pm$2.2000&0.5453$\pm$0.6270\\
Log(Luminosity)& $39.37(^{+0.00}_{-0.03})$&$39.15(^{+0.03}_{-0.05})$&$38.96(^{+0.07}_{-0.02})$&$38.89(^{+0.01}_{-0.10})$&$38.71(^{+0.07}_{-1.32})$&$38.71(^{+0.07}_{-0.21})$\\
& & & & & & \\
\hline
              &                &               &               &                 &               &          \\ 
&                &               &      B       &               &               &                  \\ 
\hline
              &                &               &               &                 &               &          \\ 
nH (10$^{22}$cm$^{-2}$) &0.002$\pm$0.002 &0.080$\pm$0.020 &0.008$\pm$0.008 &0.160$\pm$0.020 &0.008$\pm$0.008 &0.003$\pm$0.003\\
kT$_{mekal}$ (keV) &0.59$\pm$0.02   &0.38$\pm$0.17 &0.53$\pm$0.13   &0.70$\pm$0.70   & 0.44$\pm$0.12  & 0.45$\pm$0.15\\
kT$_{bremss}$ (keV) & 2.9$\pm$0.4    &5.6$\pm$1.2   &1.7$\pm$0.6     &2.5$\pm$0.3   &6.6$\pm$2.5     &4.8$\pm$1.4\\
$\chi$$^{2}$/$\nu$ & 1.18(264.51/224)& 1.44(142.41/99)& 0.67(24.15/36)&0.65(35.34/54)&1.03(47.30/46)&1.23(58.88/48)\\
F$^{unabs}_{0.3-10}$(ergs/cm$^{2}$/s) (10$^{-13}$)&2.8255$\pm$0.0103&1.8311$\pm$0.2519&1.0892$\pm$0.0819&1.0347$\pm$0.0540&0.5645$\pm$0.3267&0.5459$\pm$0.2047\\
Log(Luminosity)& $39.43(^{+0.02}_{-0.05})$& $39.24(^{+0.14}_{-0.05})$&$39.01(^{+0.01}_{-0.08})$&$38.99(^{+0.05}_{-0.11})$&$38.73(^{+0.03}_{-0.09})$&$38.71(^{+0.04}_{-0.09})$\\
& & & & & & \\
\hline
\end{tabular}
\end{minipage}
\end{table*}

\section{RGS spectra of NGC 2903}
\label{sec:rgs}

\begin{figure}
\centering
\includegraphics[width=6.5cm,angle=-90]{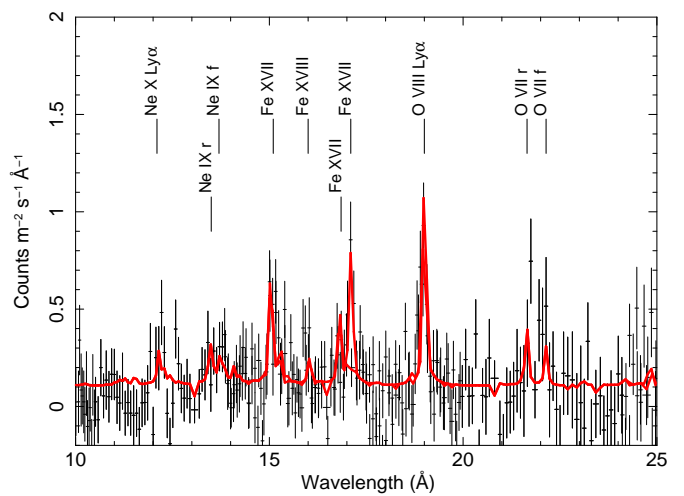}
\caption[]{RGS spectrum of NGC~2903. RGS1 and RGS2 spectra are plotted together. The most prominent emission features are labeled.}
\label{rgsplot}
\end{figure}

The Reflection Grating Spectrometer (RGS) onboard {\it XMM-Newton}
instrument was operating during the observation. The RGS data were
processed using the standard pipeline. The spectral analysis of RGS
data was carried out using the fitting package {\tt
SPEX}\footnote{http ://www.sron.nl/spex/} v2.02 \citep{kaa96}. Both
RGS1 and RGS2 were fitted together, and C-statistics was adopted as
the fitting method \citep{cash79}. Typical errors in the RGS data are
about 68\% or 1 $\sigma$. The combined first-order RGS1 and RGS2
spectra display several emission lines (see Fig.~\ref{rgsplot}). The
spectrum is dominated by a strong O\,{\sc viii} Ly$\alpha$ emission
line along with Ne\,{\sc x} Ly$\alpha$, and several Fe\,{\sc xvii}
features. The O\,{\sc vii} complex is also significantly detected,
although only the forbidden and resonance lines could be
resolved. Both O\,{\sc vii} $f$ and $r$ lines seem to be of similar
strength, which is consistent with the presence of the collisionally
ionized gas typical of starburst galaxies. The He-like triplet of
Ne\,{\sc ix} is also present, although its components are blended.

We therefore fitted the spectrum using a model for a plasma in
collisional ionization equilibrium (CIE, see {\tt SPEX} manual for
details). The spectrum was rebinned by a factor of 8 in the
7-26~\AA~range, where the majority of the emission features fall.

Even though the quality of the spectrum left little space for a
detailed quantitative analysis, we modelled the continuum with a power
law corrected for local absorption along the line of sight of the source
($N_H = 2.91 \times 10^{20}$~cm$^{- 2}$, \citealp{kal05}). The
best-fit CIE model yielded a plasma temperature of $T = 0.31 \pm
0.01$~keV and an emission measure $EM \equiv n_Hn_eV =
6.4_{-0.4}^{+0.5}\times 10^{61}$~cm$^{-3}$. Unfortunately, the
statistics of the spectrum did not allow us to test whether a
multi-temperature distribution is present or not.

The presence of the He-like triplets of O\,{\sc vii} and Ne\,{\sc ix}
in the spectrum made it possible to perform a rough line diagnostic of
the plasma. In the case of O\,{\sc vii}, the resonance $r$ and
forbidden $f$ lines are clearly detected, whereas the intercombination
$i$ line is missing or its significance is too low to be statistically
detected. The strengths of the $r$ and $f$ lines are comparable and
provide a gas ratio of $G(T_e)=(f+i)/r=1.06\pm0.52$, fully consistent
with a collisionally ionised gas. According to Fig. 7 in \citet{PD00},
this ratio for a plasma with temperature $T \sim 3.5 \times 10^6$~K
(as provided by our best-fit model) would correspond to a ratio of the
relative ionic abundances in the O\,{\sc viii}/O\,{\sc vii} ground
state population of 10, in agreement with the relative strengths of
the oxygen lines detected in the RGS spectrum.  The ratio $R(n_e)
\equiv f/i$ is far more difficult to estimate since the
intercombination line is not detected in our spectrum. Nevertheless,
that the $r$ and $f$ lines are present and of similar strength
combined with an undetectable $i$ line makes it possible to set a
lower limit to the density of $n_e \geq 10^{10}$~cm$^{-3}$ (see Fig
. 11 in \citet{PD00}. The Ne\,{\sc ix} triplet is clearly detected in
the spectrum but the individual lines are blended. In this situation
little can be said about this triplet.

From our best-fit model, we can also measure the abundances of various
elements relative to the standard set of proto-solar abundances by
\citet{l09}. We find that the abundances of oxygen, neon, and iron are
consistent with solar values ($1.0\pm0.1$, $0.85\pm0.25$, and
$1.06\pm0.08$, respectively).

\section{Conclusions}
\label{sec:concl}

We have reported on the first deeper X-ray broad-band observation of
the hot spot galaxy NGC~2903 obtained with \textsl{XMM-Newton}. Both
X-ray imaging and spectra of the spiral barred galaxy NGC~2903 were
extracted from \textsl{XMM-Newton} archival data. A total of six X-ray
sources (four of them ULX candidates) were detected in the energy
range of 0.3--10.0\,keV located within the $D_{25}$ optical disk, thus
associated with the galaxy. Three of these sources have been detected
for the first time, and one, XMM-NGC2903~X2 of luminosity higher than
10$^{39}$ erg s$^{-1}$. After fitting three different models, we were
able to estimate their luminosities, which are compatible with
binaries that contain a compact object in the form of BHs rather than
a NS.

We extracted the combined first-order RGS1 and RGS2 spectra of its
central region, which exhibit several emission lines. Both O\,{\sc
vii} $f$ and $r$ lines seem to be of similar strength, which is
consistent with the presence of collisionally ionized gas, typical of
starburst galaxies. The He-like triplet of Ne\,{\sc ix} is also
present, although its components are blended. After fitting the
spectra with a model for a plasma in collisional ionization
equilibrium (CIE) and modelling the continuum with a power-law, the
estimate of the plasma temperature was $T = 0.31 \pm 0.01$~keV and the
emission measure $EM \equiv n_Hn_eV =6.4_{-0.4}^{+0.5}\times
10^{61}$~cm$^{-3}$. From our best-fit model, we have also been able to
measure the abundances of various elements relative to the standard
set of proto-solar abundances, and found that the abundances of
oxygen, neon, and iron are consistent with solar values ($1.0\pm0.1$,
$0.85\pm0.25$, and $1.06\pm0.08$, respectively).

\begin{acknowledgements}

DPR acknowledges support from the Universidad de Ja\'en (Spain). MCG
acknowledges hospitality at {\it Departament de Astronomia i
Meteorologia (Universitat de Barcelona, Spain)}.  The Space Research
Organization of The Netherlands is supported financially by NWO, the
Netherlands Organization for Scientific Research.

\end{acknowledgements}

\end{document}